WHITEPAPER

# First International Workshop on Serverless Computing (WoSC) 2017

# Report from workshop and panel on the Status of Serverless Computing and Function-as-a-Service (FaaS) in Industry and Research

Geoffrey C. Fox (Indiana University)

Vatche Ishakian (Bentley University)

Vinod Muthusamy (IBM)

Aleksander Slominski (IBM)

This whitepaper summarizes issues raised during the First International Workshop on Serverless Computing (WoSC) 2017 [1] held June 5th 2017 and especially in the panel [2–5] and associated discussion that concluded the workshop. We also include comments from the keynote [6] and submitted papers [7–10]. A glossary at the end (section 8) defines many technical terms used in this report.

**Panel participants**: Geoffrey C. Fox (Indiana University), Rodric Rabbah (IBM), Garrett McGrath (University of Notre Dame), Edward Oakes (University of Wisconsin-Madison), Ryan Chard (Argonne National Laboratory), and Ali Kanso (IBM)

## 1 Introduction

Panel participants were asked to provide a short presentation for one of suggested topics

- Describe current state of field in terms of technology and adoption

- Argue that serverless computing is nothing new and point out the relevant literature and past achievements
- Take the position that serverless computing is fundamentally different and requires revisiting common assumptions.
- Discuss challenging real-world problems that could be research issues.
- Outline the definition and scope of serverless computing platforms.
- Propose a benchmark to compare serverless platforms.
- Suggest a timeline for evolution of technology and adoption for area

The panel and workshop presentations are linked from the workshop website.

In this whitepaper we will only summarize and emphasize the themes that were raised during panel and workshop - the detailed notes are available as a separate document.

We believe that serverless computing [11] is not only an exciting platform for researchers to explore but also for academia to use. There are upcoming changes in leading cloud analytics platforms to become more serverless (for example Spark [12]) and some experiments to use serverless directly as runtime for analytics (for example [13]).

## 2 Basic Definition of Serverless and FaaS

Serverless evolved over time as shown in Fig. 1. The beginning of usage of the term 'serverless' can be traced to its original meaning of not using servers and typically was applied to peer-to-peer (P2P) software or client side only solutions [14,15]. In the cloud context, serverless started to mean that developers do not need to worry about servers and in particular just uses SaaS platforms or services such as Google App

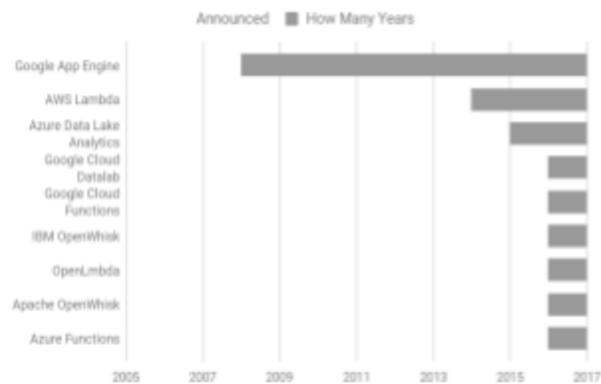

Fig. 1: Timeline of different serverless services.

Engine [16–18]. The latest serverless solutions are really server-hidden and built to host functions and hide that the functions runs on servers or how scaling is done. The functions may be part of a service (for example Azure Data Lake Analytics  or Google Cloud Datalab) or offered as an independent service called Function-as-a-Service or FaaS. Note that unlike SaaS or PaaS that are always running, but scale on-demand, serverless workloads run on-demand, and consequently, scale on-demand. Summarizing this, we see that the same term serverless is being used to describe related but different concepts.

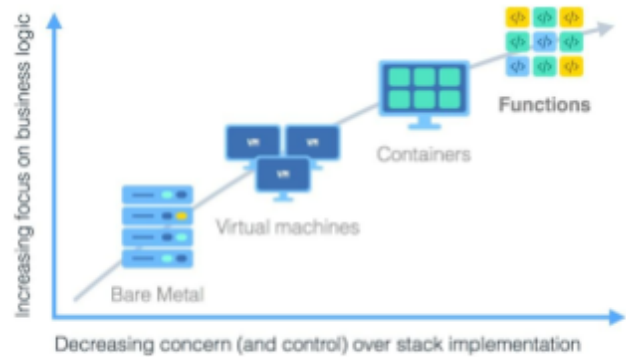

Fig. 2: Progression of System Deployment Options

From the IBM tutorial at workshop [19,20], we find their definition of FaaS and Serverless as

- A cloud-native platform
- For short-running, stateless computation
- And event-driven applications
- which scales up and down instantly and automatically
- And charges for actual usage at a millisecond granularity

Fig. 3: Comparison of 3 IaaS Deployment Options

Fig. 2 shows the evolution of Infrastructure or IaaS from an old data center model with explicit servers to serverless which was described by Barga in his keynote [6] with the tag line that "No server is easier to manage than no server". More details of this evolution are given in Figs. 3 and 4. Hiding the server infrastructure as in Serverless is coming to attention just as public clouds offer an increasingly rich variety of instances with compute, memory, accelerator, and I/O choices that offer amazing functionality but at increasing complexity.

Fig. 4: Comparison of serverless and traditional data center models in 4 key deployment and cost areas

Fig. 5 summarizes some of areas where today serverless may excel or have limitations. However discussion at the meeting suggested that this characterization could change. For example today FaaS, Event driven computing, stateless, and short running are all associated with serverless. However we can expect these important ideas to evolve independently and

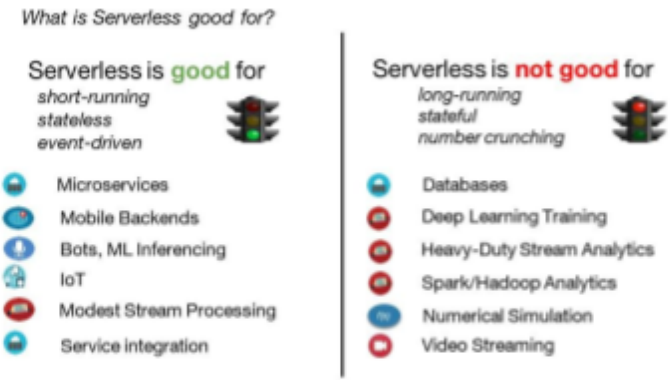

Fig. 5: Pluses and Potential minuses of Serverless Computing

not be tied closely together. For example, event driven FaaS could support long running jobs and/or be offered on explicit IaaS. Contrastly serverless ideas (hiding the details of server deployment) could be used on many different cloud computing scenarios. In the keynote, Barga described Amazon Lambda which is their event driven computing model underlying their serverless offering. The Lambda homepage [21] describes serverless FaaS well:

*"AWS Lambda lets you run code without provisioning or managing servers. You pay only for the compute time you consume - there is no charge when your code is not running. With Lambda, you can run code for virtually any type of application or backend service - all with zero administration. Just upload your code and Lambda takes care of everything required to run and scale your code with high availability. You can set up your code to automatically trigger from other AWS services or call it directly from any web or mobile app."*

Examples of the breadth of serverless included the PyWren MapReduce based on FaaS [13] and the importance of the event driven computing model to edge computing (see section

5) -- identified as joining serverless as two separate drivers of next generation cloud computing.

# 3 Comments on Serverless and FaaS Technologies: The State of Serverless Computing

## 3.1 The definition compared to current practice in serviceless and FaaS

One issue that was raised often was the definition of serverless and Function-as-a-Service (FaaS) already brought up in section 2. The serverless manifesto poses this well [22]. During his workshop keynote [6], Roger Barga defined serverless as the next stage of in an evolution of cloud computing from Grid to IaaS Cloud to PaaS/SaaS to serverless FaaS, where developers can build services without worrying about servers; both event driven and stateless did not seem essential -- just common features. He used the graphic shown in Fig. 6. That definition is much broader than small stateless functions or FaaS. In contrast the IBM tutorial [20], defines serverless as Small Stateless Functions as a Service, which fits the current state of Apache OpenWhisk which was a centerpiece of their contribution. Note small functions naturally fit the growing use of microservices. Note the smallness of functions with short running times ("Kill after 5 minutes" and "transient residency") is important on the provider-side as it allows low costing of FaaS which is used to fill the load between larger jobs on a cloud infrastructure. Still this feature including its important lower cost, could evolve to just one of several serverless offerings. We can discuss the relation between Apache Storm (Heron or Amazon Kinesis) compared to Apache OpenWhisk (or Amazon Lambda). How does the dataflow model in Storm (or Spark and Flink) relate to FaaS? Eventually FaaS could be an implementation. We can also compare SaaS with FaaS; is the latter an advanced subset of former? And in general what are the intersections and overlaps between traditional serverless (where servers literally are not used instead code is running peer-to-peer to provide services such as storage, messaging, etc.) and event-driven computing (see Fig. 7)? And is FaaS limited to event-driven computing ?   Note Figs. 6 and 7 are inconsistent in detail with the nested classifications of Fig. 6 being relaxed in Fig. 7; this just reflects the typical confusion in an emerging field.

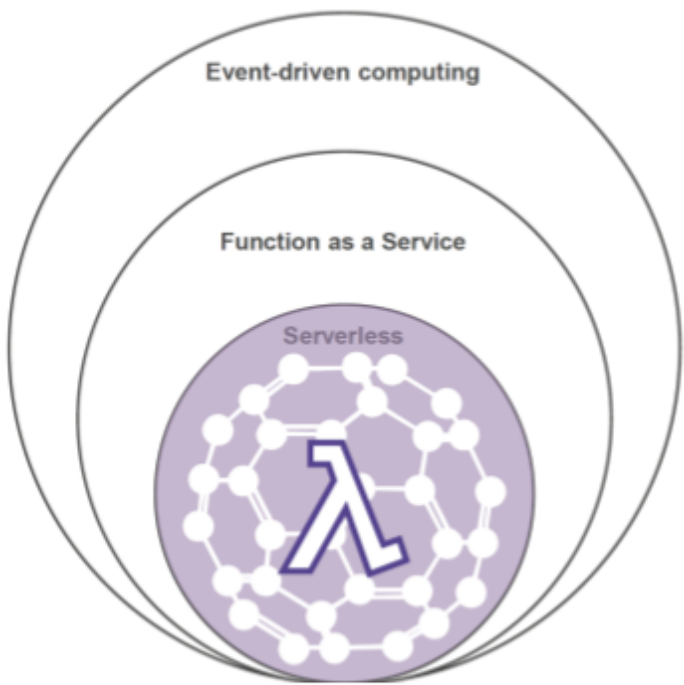

Fig. 6: Event Driven Computing, FaaS and serverless in AWS Ecosystem

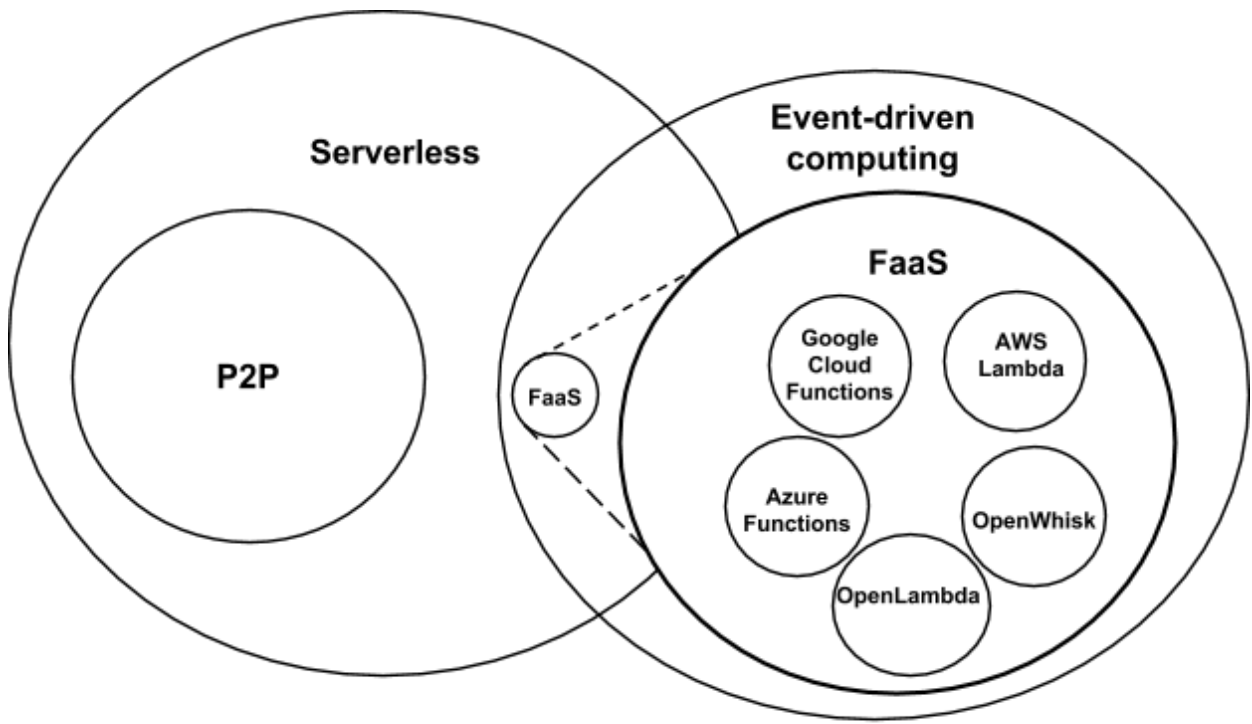

Fig. 7: One view on the relation between serverless, P2P, event-driven computing, and FaaS.

This discussion leads to natural questions such as why serverless is a good name when you need to explain what it is? And why not just call it function computing or FaaS, if it is all about stateless functions? Is serverless just a specialized option or is it good for almost everything?

The Cloud Native Computing Foundation CNCF Serverless Working Group is exploring the intersection of cloud native and serverless technology and their web resource [23] has a substantial accumulation of useful information on serverless and FaaS.

The increasing importance of Serverless computing is illustrated by the appearance of the term "Serverless PaaS" which is "on the rise" in the 2017 Hype Technologies Report [24] from Gartner.

## 3.2 What is new about Serverless?

Rodric Rabbah brought forward the recent example of the FCC website that collapsed when it was unable to handle comments about net neutrality. That is good example where serverless could be making an immediate real difference - if the FCC used a serverless platform that would have a better chance to handle the scale of traffic generated. Trying to decide how many servers to deploy and then maintain their scaling is hard job and unless substantial expertise is available in-house it is easy to make mistakes. This example brings up the support of elasticity and cloud-bursting to reach larger capacity sites; scheduling technology needs to be improved to support this.

What is also making serverless attractive is a cloud offering of an ecosystem of supporting middleware and artificial intelligence services that integrate seamlessly with the serverless platform to enable natural language processing, image recognition, manage state, record and monitor logs, send alerts, trigger events, or perform authentication and authorization. The use of such services not only present another revenue stream for the cloud provider, but also enables application dependency on the provider's ecosystem and vendor lock-in.

Serverless builds upon technologies that have been subject of previous research in different computing domains, what is particularly new about serverless? Is Serverless be all and end all of new technologies? What is the real cost of Serverless?

## 3.3 Is Serverless Necessarily Stateless?

The stateless or stateful aspect of serverless produced much discussion. Storing state external to a "stateless" FaaS could enable many important applications and allow big datasets to trigger multiple microservice-based FaaS invocations. Here we can look at AWS Step Functions which can orchestrate a workflow of multiple microservices while RDD in Spark can store state in an external entity that can easily be accessed by using an in-memory database. Note the manifesto [22] SLE assertion that in serverless: "permanent Storage Lives Elsewhere".

## 3.4 Provider Side view of Serverless

This was discussed in McGrath's panel presentation [5] with serverless computing allowing providers to understand customer applications and to deliver value based on this information. Applications declare behavior such as the triggering events and one can also predict behavior -- perhaps with machine learning from logs. The serverless fine grained programming model gives the provider more flexibility to schedule/optimize. There is perhaps a relation to JIT compilers here.

There are mutual economic pressures as Cloud providers need to cost-compete by running datacenters more efficiently (utilization, energy-efficiency) while Cloud customers seek to reduce cost by minimizing resource waste. Both can be satisfied by better matching of application needs to allocated services. Serverless computing is a large step forward but we're not there yet as we ask for "Never pay for idle, or for wait" [25] as time spent waiting on network (function executions or otherwise) is wasted by both provider and customer. Here the billing model of serverless is questioned. The simple view is that one only pays for what one uses but network delay can lead to billing for unused time.

## 3.5 Can serverless work for longer running tasks?

We discussed the compatibility of serverless with long running compute tasks with different aspects of this being illuminated by the panelists. Long running jobs are of course well known in High Performance Computing (HPC) with sophisticated scheduling based on user time estimates: serverless workloads today are very short lived but maybe in the future will be longer as in HPC. The provider will need to provide a service level agreement (SLA) and long running tasks give the provider less flexibility in scheduling and more difficulty in cost-effective SLA's. Of course, serverless gives the illusion of unlimited resources and one "just" needs to realize this. One possibility is to handoff long running jobs to a different container service. Alternatively, AWS Step Functions let you manage long running flows by combining multiple (small) FaaS invocations.

This question forces one to address the different facets of Serverless independently: hidden (from user) IaaS, event-based, edge workflows, streaming data, dataflow, micro (time, size) services. If long running jobs are allowed, you will presumably need checkpointing.

## 3.6 Standards

The question of standards was discussed with the clear goal of supporting easy movement of business logic between different serverless platforms and prevent vendor lock-in. There are currently no directly applicable standards although it is early days to set standards for a capability that is still being defined. Further we know that AWS is the market leader of the field and may not have a clear motivation to develop standards other than the de facto standard -- their technology. It was noted that a rationale for open sourcing OpenWhisk is to build a community from which standards can be developed. Further CNCF has a very relevant working group [15]. Again at this early stage, many smaller players can still upset the market leader.

Messaging standards, including the machine to machine light-weight pub-sub system MQTT [18], could relevant while the importance of the generally used Robot Operating System [19] could lead to standards.

## 3.7 Programming models

We discussed possible programming models (reactive programming, logic programming, functional, etc.) that could be appropriate to address FaaS including the problem of moving compute around. Of course as with standards, we are right at the beginning and we can expect a lot of opportunities for innovation in programming languages and runtime. Although event-based programming is not totally new, the use in the datacenter is a new context, while IoT devices need to worry about energy usage. The intersection of FaaS and traditional Big Data programming environments such as Spark, Flink, Hadoop, Storm and Heron is discussed in [26,27].

## 3.8 Are there any cons to Serverless and FaaS?

There was a lengthy discussion of the possible negatives and difficulties with FaaS and serverless. At the highest level there was concern that users (industry) were chasing the latest fad (in this case serverless) without consideration of the soundness of the approach. For example, there are still significant challenges in using OpenStack and Docker at scale. In latter case, OpenWhisk uses Docker at an unprecedented scale and has uncovered many concurrency bugs.

Concerns were expressed about maintaining the (attractive) pricing model for Serverless. This is important for keeping cost down for intermittent streaming applications. Note that as one uses Serverless for more complex applications, the provider will get additional funds from the incidental activities such as traditional storage (save state), a supporting ecosystem of available provider functionality, and computing in the cloud at the expense of vendor lock-in. Also current (lack of) SLA for serverless may make it unattractive for latency sensitive applications in Government, Healthcare, and Banking. Serverless will not handle 911 in the near future or until SLA's are addressed seriously. In this case, one might be forced to doing FaaS oneself in a private cloud -- i.e. In fact worrying in detail about the IaaS that you tried to avoid.  A different view was expressed that this is not really a con;

serverless promotes separation of concerns between the application logic and the runtime. Today the runtime is typically in the cloud, but it could be in-house as well.

The panel discussed using Platform as a Service PaaS instead of FaaS. PaaS is compatible with scaling up the servers as needed to meet demand. For PaaS, the scaling is reactive and not deterministic as for FaaS. Further, you still need to manage the workflow and minimum number of instances for PaaS.

A comparison was made with networking with an analogy drawn between FaaS and network packet switching with both multiplexing demand. QoS is difficult in both FaaS and network packet switching with latter compared to circuit switching.

### 3.9 Current Serverless Systems

The workshop was not aimed at a comprehensive survey of existing serverless technologies but it certainly did cover the current technology to some extent. Notably the IBM Tutorial [20] gave a thorough discussion of what is now Apache OpenWhisk. The keynote from Amazon [6] naturally covered AWS technologies; important as they are the current commercial leader. As well as AWS Lambda and Kinesis, Barga covered Greengrass for IoT and X-Ray for debugging.

The Notre Dame paper [9] described their new serverless system built around Docker on Azure with Windows. They also compared this with Google, AWS. OpenWhisk, and Azure serverless systems. The performance results seemed quite erratic in this early stage of the field. This paper defines a benchmark and here we certainly need community development.

The Wisconsin paper [10] was mainly based on the Pipsqueak python packaging application but the open source OpenLambda technology was the environment used.

The value of Google Firebase as a serverless IoT tool was emphasized.

# 4 Use Cases for Serverless and FaaS

Of course the future of Serverless and FaaS will critically depend the application drivers and the breadth of user cases is driving a lot of the current interest in the field [13]. Amazon Alexa like chatbots are another example of that interest[28]. The event-based model is familiar from previous work such as CORBA on distributed object technology with RMI (Remote Method Invocation) or RPC (Remote Procedure Call) implementing FaaS. Rather old examples of this include "optimization on demand" NEOS [29] and the DoD high level architecture HLA implementation of distributed simulation [30]. NetSolve and GridSolve [31] represent the Grid community approach to RPC.

One can also argue that the cloud provider can influence the use cases for Serverless. The more self awareness (through monitoring) the cloud has (e.g. traffic patterns, resource utilization, data transfer size/frequency, …), the more triggers it can offer to its customers and the more triggers the customer have, the more functions they can write to react to those triggers. Serverless is a declarative policy-based approach such that the more triggers we have, the richer the policies can be.

## 4.1 What are established use cases for serverless?

One major use case motivation as stressed by Barga is user convenience; they do not want to worry about complex IaaS. A more specific feature is the automatic elastic scaling as is needed in many e-commerce applications such as ticket sales with surges in popularity. A broad use case is support of edge computing described in section 5 and in the following we discuss use-cases covered in papers and presentations.

Barga's keynote [6] discussed 6 classes of use-cases: web applications, backends including IoT (section 5), Big Data, Chatbots, Amazon Alexa and IT Automation. Under Big Data, Barga mentioned PyWren with 600 concurrent functions; he challenged the audience to explore more sophisticated MapReduce applications. Image thumbnail production; streaming social media data analysis in Kinesis; data warehouse ETL transformations; e-commerce recommendations; financial monitoring were discussed. Barga noted that Thomson

Reuters processes 4,000 requests per second and Expedia 1.2 billion requests per month on Lambda. The video hosting company Vevo handles spikes of a factor 80 using serverless elasticity.

The Wisconsin Pipsqueak paper [10] describes an interesting application to have a large number of Python library functions available for serverless FaaS. This was achieved with a sleeping Python interpreter and the package stored in memory and SSD. The IBM paper [7] goes through a use case where OpenWhisk is used to process results of Vulnerability scans on Docker containers managed by Kubernetes. The results of the scan posted on a policy endpoint to be processed by FaaS.

## 4.2 Can serverless help with scientific research?

The panel considered that serverless and FaaS although currently explored in business, do have major importance for science and engineering research. For example there are many scientific Instruments gathering data with custom Laboratory management systems that could be unified to advantage with FaaS. This is related to applications discussed in section 5 and has been extensively in recent workshops [32,33] on streaming data for science. The latter raised interesting questions about the functionality of systems like Apache Storm for science experiments; these typically have events such as huge images that are larger than those seen commercially. The issues of reproducibility, scalability, and cost need to be explored for science use cases.

One of the presented papers [8] discussed a science data management use case of monitoring a HPC storage workload (with over 3 million events/day). The Ripple system implements a IFTTT (if this then that) model with "that" implemented on AWS Lambda and using file system event detection for the "this" with Python Watchdog and the Globus Transfer API. Applications to astronomy and light source data analysis are being investigated.

## 5 Edge Computing: A Key Driver for Serverless and FaaS

There is a natural relevance of FaaS and edge computing as latter is inevitably built around events shared between device and fog; fog and cloud [26]. In fact this link between serverless and edge computing was an important take-away from the workshop. This edge-cloud integration can be implemented [34] with Apache Storm (AWS Kinesis) linked to Apache OpenWhisk (AWS Lambda). It was stressed that we are not proposing to move computing to the edge but rather to integrate the edge with the cloud. In fact data centers are getting larger not smaller and we are not moving back to a very distributed core computing model except for the case where we need to link datacenters to activities at the edge. Content Delivery Networks, multiplayer games, smart homes [35] and autonomous vehicles are current important examples, where the latter cases were obviously very hot with the CES show in Las Vegas January 2017 full of such startups.
iRobot use of Lambda and AWS Step Functions for Image recognition was described by Barga [6] as an example of inherently distributed serverless application. Barga further discussed AWS Greengrass [36] extending Lambda to a common cloud-device environment with interesting quote *"Amazon expects that the majority of on-premises hardware will soon be IoT devices as enterprises move their servers into the cloud. "* AWS Snowball edge storage and compute runs this Lambda@Edge software. Google Firebase is a related product.

## 6 Future: What are low hanging fruits for serverless?

The panelists were asked to discuss a timeline and topics for the evolution of the technology and a discussion of its adoption by users. The suggestions varied from wide ranging dreams to detailed nuts and bolts.

Optimistically it was predicted that FaaS will be applied to general purpose computing and it will grow in capability and limitations such as the "5 minute kill limit" will disappear. It will be great for end developers as they will not need to know scaling and distributed computing. A hot research topic will be its use for parallel programming which is Barga's challenge to extend the MapReduce use of FaaS. It will be applied to batch processing and used to reach exascale on supercomputers. Scientific notebooks need to be integrated with FaaS. FaaS could further help users by making libraries easier to use as one needn't put library routines in one's code; just invoke them as FaaS.

At a more detailed level, debugging was identified as a near term critical problem where we need to be able to test locally and then deploy on the edge and the cloud.  The debugger itself should be serverless and support live breakpoints and replay. We can adopt a test-driven development with unit tests.

Performance is an important issue although not the only one -- usability is for example a key feature of serverless. More generally, we need to define evaluation metrics [9]. Unikernels are an attractive technology for serverless. There are also security concerns to be addressed; does one need more than a container for the function and how should events be made secure?

The billing issues brought up in section 3.4, need to be studied and understood how much of the delays and overheads are inevitable. It was noted that in AWS Step Functions, one decouples the billing of the functions from the coordination of the composition.

There was substantial discussion about the programming model and runtime. For runtime, load balancing (handling communication and computing) and scheduling were identified. Note the runtime is a provider point of view (allowing magic behind the scenes) and the programming model the concern of users. Data-locality needs to built into the runtime. The programming model and runtime need to support key features of serverless: event driven, hidden servers for users, fine grained billing, implicit distribution, low latency. Analogously to the Java Virtual Machine JVM, serverless could become a common runtime for multiple programming models. The fine grain nature of FaaS allows more optimizations than those conventionally allowed; this needs research. The runtime research needs to understand what SLA's are needed and what can be supported.

The identification of common patterns for FaaS is important. This would be coupled to study of function compositions. Related to composition, we can ask where the "main program" is located -- does it run (as in some workflow systems) outside the FaaS environment. Serverless is right at its start; just as Spark improved on the original MapReduce, we need the next generation FaaS (which is in fact compatible with Spark, Flink and Heron!)

Both serverless technologies and their evaluation are immature. We need to develop benchmarks covering both edge computing and other use cases. This workshop attempts to address another need; the development of a serverless developer community.

# 7 Conclusion

To some, serverless and FaaS are the next generation of computing supporting centralized and edge computing with a common event-driven programming model. Conversely the drivers of cloud computing are Edge Computing and Serverless. One often discusses distributed / edge computing versus centralized approaches and wonders how we move back and forth; the answer is clear -- we have both intrinsically intertwined.  Serverless will build the long dreamed infinite limitless computing fabric.

This white paper aims to capture the current state of serverless and FaaS and hopefully inspire a broader community to become involved.

# 8 Glossary

**Apache/IBM OpenWhisk:** Apache OpenWhisk (Incubating) is a serverless, open source cloud platform that executes functions in response to events at any scale http://openwhisk.incubator.apache.org/. It builds on IBM Bluemix project https://developer.ibm.com/openwhisk/.

**Apache Storm and Heron:** Open source programming and execution on the cloud for data streaming. Systems originally developed by Twitter with Heron improving Storm with same API. http://storm.apache.org/   https://twitter.github.io/heron/

**AWS Kinesis:** collect, process, and analyze real-time, streaming data in a similar fashion to Apache Storm https://aws.amazon.com/kinesis

**AWS Greengrass:** Amazon Lambda supporting local compute, messaging, data caching, and sync capabilities on a device at the edge [36]   https://aws.amazon.com/greengrass/

**AWS Lambda:** Event-based computing FaaS from Amazon [21]

**AWS Step Functions:** lightweight orchestration of Amazon Lambda Functions as distributed applications using visual workflows.https://aws.amazon.com/step-functions/

**AWS X-Ray:** analyzes and debugs distributed applications, such as those using microservices and Amazon Lambda  https://aws.amazon.com/xray/.

**Azure Functions:** Implementation of serverless FaaS on Azure. https://docs.microsoft.com/en-us/azure/azure-functions/functions-overview

**Cloud Native:** applications are designed to run on clouds as preferred host, exploiting concepts such as containers, microservices, elasticity and serverless https://www.cncf.io/ [37]

**Content Delivery Network CDN:** is a geographically distributed network of proxy servers that distribute information from data centers to spatially distributed users with high availability and high performance. CDNs serve a large fraction of the Internet content toda https://en.wikipedia.org/wiki/Content_delivery_network

**Dataflow:** describes a range of computing ideas but here refers to an execution graph defined by data flowing between nodes as seen in Apache Storm Spark and Flink.

**Docker:** Open Source container technology for Linux and Windows supporting operating-system-level virtualization  https://en.wikipedia.org/wiki/Docker_(software)

**Edge Computing:** The processing associated with the Internet of Things IoT and including local computing resources, often termed Fog computing, devoted to give local low-latency support to IoT devices. https://en.wikipedia.org/wiki/Fog_computing

**Function as a Service FaaS:** Event based functions typically executed on serverless infrastructure and described in section 2 of report

**Funktion:** Open source event driven lambda style programming model on top of Kubernetes. https://github.com/funktionio/funktion

**Globus Transfer:** cloud-controlled secure high-performance data transfers based on advanced FTP https://www.globus.org/data-transfer

**Google Firebase:** A mobile development platform linking to the cloud and exploiting serverless computing Google Functions to process events. https://firebase.google.com/

**Google Functions:** FaaS provided on Google clouds https://cloud.google.com/functions/

**GridSolve:** implemented as Netsolve, is an RPC based client/agent/server system that allows one to remotely access computing functions as a service [31].

**High Performance Computing HPC**: a community and an approach built to support the largest scale computational science, especially numerical simulations. Typically uses supercomputers and achieves very efficient batch scheduled execution. A prominent use of HPC in Big Data is the training of deep learning networks.

**Infrastructure as a Service IaaS:** makes servers explicit for users of cloud computing but abstracts away the details of infrastructure like physical computing resources, location, data partitioning, scaling, security, backup etc.

**Kubeless:** Kubernetes-native serverless framework https://github.com/kubeless/kubeless

**Kubernetes:** Open-source platform for automating deployment, scaling, and operations of application containers such as Docker across clusters of hosts https://kubernetes.io/docs/concepts/overview/what-is-kubernetes/.

**Kubernetes Fission:** Serverless Functions as a Service for Kubernetes developed by Platform9 http://blog.kubernetes.io/2017/01/fission-serverless-functions-as-service-for-kubernetes.html

**Microsoft Logic Apps:** provides a visual interface to specify workflows of connected applications and triggers in the cloud. https://docs.microsoft.com/en-us/azure/logic-apps/

**Microservice:** service-oriented architecture (SOA) style that structures an application as a collection of loosely coupled fine-grained services communicating by lightweight protocols. https://en.wikipedia.org/wiki/Microservices

**OpenLambda:** Open-source serverless computing platform. https://open-lambda.org

**Platform as a Service PaaS:** provides a cloud development environment (middleware) with details of underlying resources often hidden. https://en.wikipedia.org/wiki/Cloud_computing

**Pipsqueak:** Serverless package support from the University of Wisconsin - Madison [10]

**PyWren:** Python MapReduce with stateless maps running under Amazon Lambda [13]

**Ripple:** Science event based FaaS application for data management [8]

**Serverless:** discussed in this whitepaper as a server hidden cloud computing execution model where provider dynamically manages the allocation of machine resources, and bills on use rather than on pre-purchased units of capacity. https://en.wikipedia.org/wiki/Serverless_computing

**Software as a Service SaaS:** is a cloud computing usage model where providers install and operate application software in the cloud, which is accessed by cloud users.

**Unikernels:** are specialised, small specialized high performance single address space machine images constructed by using operating systems such as MirageOS [38] built as a library of system capabilities. https://en.wikipedia.org/wiki/Unikernel

## 9 Acknowledgements

Geoffrey Fox acknowledges support from the Indiana University Precision Health Initiative and NSF CIF21 DIBBS 1443054. We also acknowledge the contributions of both the panel members and participants in the workshop which were essential to the ideas described here.